\documentclass[5p,number]{elsarticle}

\usepackage{graphicx}
\usepackage{scrextend}
\usepackage{dcolumn}
\usepackage{bm}
\usepackage[strings]{underscore}
\usepackage{textcomp}
\usepackage{mathtools}
\usepackage{float}
\usepackage{amsmath}
\usepackage{pifont}
\usepackage{natbib}
\usepackage{geometry}
\usepackage{fleqn}
\usepackage{color}
\addtokomafont{labelinglabel}{\sffamily}
\biboptions{sort&compress}
\graphicspath{{./eps/}}

\journal{Journal of Alloys and Compounds}

\begin{document}
\begin{sloppypar} 

\begin{frontmatter}

\title{Electronic structure of CeCo$_{1-x}$Fe$_x$Ge$_3$ studied by X-ray photoelectron spectroscopy and first-principles calculations}

\author[1]{P. Skokowski\corref{cor1}}
\ead{przemyslaw.skokowski@ifmpan.poznan.pl}
\cortext[cor1]{Corresponding author}

\author[1,2]{K. Synoradzki}

\author[1]{M. Werwi{\'n}ski}

\author[3]{A. Bajorek}

\author[3]{G. Che{\l}kowska}

\author[1]{T. Toli{\'n}ski}

\address[1]{Institute of Molecular Physics, Polish Academy of Sciences, Smoluchowskiego 17, 60-179 Pozna{\'n}, Poland}
\address[2]{Institute of Low Temperature and Structure Research, Polish Academy of Sciences, Ok{\'o}lna 2, 50-422 Wroc{\l}aw, Poland}
\address[3]{Institute of Physics, Silesian University, Uniwersytecka 4, 40-007 Katowice, Poland}
\date{}

\begin{abstract}

%
A transformation between the magnetically ordered CeCoGe$_3$ and heavy fermion CeFeGe$_3$ is isostructural but not isoelectronic, therefore the characterization of the electronic structure of the CeCo$_{1-x}$Fe$_x$Ge$_3$ series is of special importance. 
%
%
%
We report both the experimental investigation by the X-ray photoelectron spectroscopy (XPS) measurements and the first-principles calculations within the full-potential local-orbital (FPLO) scheme based on the density functional theory.
%
Experimentally, we investigate mainly the valence band and the Ce $3d$ spectra, both giving the possibility to conclude about the level of the $f$ states localization.
Computationally, we investigate the most characteristic CeCoGe$_3$, CeCo$_{0.5}$Fe$_{0.5}$Ge$_3$, and CeFeGe$_3$ compositions.
Based on the electronic band structure results we calculate the X-ray photoelectron spectra of the valence band.
We consider the effects of the spin-orbit coupling and intra-atomic Hubbard U repulsion.
Furthermore, we discuss the charge distribution and occupation of valence-band orbitals.
%
%
The experimentally observed evolution of the 3$d$ band with the Fe concentration is related to the decrease of the number of electrons and reduction of the 3$d$ photoionization cross-sections. 
Calculations indicate that charge is transfered mainly from the Ce to Ge sites and the bondings are formed mainly by the Ce 5$d$, Fe/Co 3$d$, and Ge 4$p$ and 4$s$ orbitals.

\end{abstract}


\end{frontmatter}

\section{\label{intr}Introduction}

In a classical phase transition, the system properties are determined by thermal fluctuations. If the temperature of the magnetic phase transition is reduced to a value close to 0 K, quantum fluctuations start to play a key role. If this transition is continuous and is a result of competition of the magnetic interactions, e.g. of RKKY type (Ruderman-Kittel-Kasuya-Yosida) and the Kondo interaction, one says about a quantum critical point (QCP) at 0 K. However, the effect of strong quantum fluctuations at QCP also influences the physical properties at temperatures above 0 K $-$ e.g. superconductivity or non-Fermi liquid (NFL) behaviour may appear \cite{lohneysen2007fermi,continentinoquantum2005}.

Such phenomena are often observed in Ce-based intermetallics owing to the presence of the valence instability and the hybridization effects. The unusual behaviours include the creation of the heavy fermion (HF) state, fluctuating valence, as well as superconductivity and NFL.
If the system is close to QCP, it can be approached by application of external pressure or magnetic field, alternatively one can use chemical pressure, i.e. controllable substitutions in the compound. The latter usually means starting with a magnetically ordered compound and damping its ordering temperature down to 0 K by isostructural substitutions. We have recently extended such studies for CeCo$_{1-x}$Fe$_x$Ge$_3$ series, which had been previously suggested to exhibit QCP for $x$ in the range 0.5$-$0.6 \cite{demedeirosphase2001}. In this system the transformation occurs between CeCoGe$_3$, which is characterized by three antiferromagnetic phase transitions ($T_{N1} = 21$ K, $T_{N2} = 12$ K, and $T_{N3} = 8$ K \cite{pecharskyunusual1993,thamizhavelunique2005,budkomagnetoresistance1999,kanekomultistep2009,smidmanneutron2013}), and heavy fermion paramagnetic CeFeGe$_3$ compound \cite{yamamotonew1994,yamamotocefege1995}. Both materials belong to the Ce$TX_3$ ($T =\ $transition metal and $X =\ $Si, Ge, Al) 
family and they crystallize in the tetragonal non-centrosymmetric BaNiSn$_3$-type structure (space group $I$4$mm$, No. 107), see Fig.~\ref{fig1}(a). The main reason for the growing interest in the Ce$TX_3$ phases is related to the presence of the superconductivity \cite{yanase2007magnetic,kawaimagnetic2008,bauer2012non,smidmanneutron2013}. While the influence of the external pressure on the Ce$TX_3$ systems has been quite often investigated, the studies on the effects of the chemical composition change are rather rare in the literature \cite{demedeirosphase2001,eom1998suppression,continentino2001anisotropic,rosch2005comparison,xia2018tuning}. Our recent studies have concerned the signatures of the NFL observed by measurements of the electrical resistivity, magnetoresistance, specific heat, and thermoelectric power \cite{skokowski2018complex}.

It is important to notice that the transformation between CeCoGe$_3$ and CeFeGe$_3$ is isostructural but not isoelectronic, as obviously follows from the electronic structure of Co and Fe. Therefore, in this paper we present results of the studies of the electronic structure both experimentally, by X-ray photoelectron spectroscopy (XPS) measurements, and theoretically, by first-principles calculations using  generalized gradient approximation (GGA)~\cite{perdew1996generalized} and including the interaction term U on Ce 4$f$ orbitals~\cite{czyzyk1994local}. The aim is to survey the valence behavior and the partial contributions to the density of states for various Fe concentrations $x$ in the CeCo$_{1-x}$Fe$_x$Ge$_3$ series.

The first-principles calculations are based on the fundamental laws of physics and do not require the use of any fitting parameters acquired from experiment.
For solids, the first-principles calculations describe the interacting electrons and nuclei within the framework of quantum mechanics.
The appearing numerical complexity of the many-body problem of interacting electrons can be overcome by application of the density functional theory (DFT). 
However, the DFT can be applied only with the use of approximations of an exchange-correlation functional like, for example, the local-density approximation (LDA) or generalized-gradient approximation (GGA), in which the exchange-correlation energy depends on the electronic density.
The considered solids are modeled based on a crystallographic unit cell (defined by lattice parameters and atomic positions) with the three-dimensional periodic boundary conditions.
In this work the first-principles calculations are performed using the FPLO code \cite{koepernik1999full}.
The brief summary of DFT and FPLO can be found, for example, in a book chapter authored by Eschrig~\cite{Eschrig2004}.
Among the DFT codes FPLO is characterized by high efficiency and at the same time it provides the most accurate numerical results \cite{lejaeghere2016reproducibility}.
The FPLO owes its high precision to, among others, an application of full-potential method providing no shape approximation to the crystal potential and to an expansion of the extended states in terms of localized atomic-like basis orbitals \cite{Eschrig2004,koepernik1999full}.
The use of full potential is particularly important because the accuracy of results for 4$f$ systems depends on the quality of the potential \cite{koepernik1999full}.
Furthermore, the application of full relativistic version of the FPLO code (including spin-orbit coupling) improves the description of the 4$f$ electrons characterized by a large spin-orbit coupling parameter.
The terminal compositions, CeCoGe$_3$ and CeFeGe$_3$, have been studied from first principles previously~\cite{jeong2007electronic, chigo2006lsda}.
For CeCoGe$_3$ Jeong has also used the FPLO code~\cite{jeong2007electronic}, however our approach goes a little further taking into account spin-orbit coupling and intra-atomic Hubbard U repulsion term within a single model.
Furthermore, we use gradient functional instead LDA.
For CeFeGe$_3$ Chigo-Anota \textit{et al.}~\cite{chigo2006lsda} have used the tight binding - linear muffin-tin orbitals - atomic sphere approximation code (TB-LMTO-ASA), however considering no spin-orbit coupling.

\section{\label{exp}Methods}
\subsection{\label{rxps}Experimental}
The polycrystalline samples of the CeCo$_{1-x}$Fe$_x$Ge$_3$ series were synthesized in the induction furnace. To ensure homogeneity they were turned upside-down and remelted several times. In addition, the samples were wrapped in the Ta foil, placed into quartz tubes and annealed at $750^\circ {\rm C}$ for $120 \ {\rm h}$. The weight losses of all samples after melting were less than 0.5$\%$ of the total mass. X-ray diffraction showed that all the studied compounds are isostructural, single phase, and crystallize in the tetragonal BaNiSn$_3$-type structure devoid of the inversion symmetry ($I$4$mm$ space group).

The XPS measurements were performed with the use of PHI 5700/660 Physical Electronics spectrometer. The spectra were analyzed at room temperature using monochromatized Al K$\alpha$ radiation (1486.6 eV). The samples were fractured and measured in vacuum of 10$^{-10}$ Torr.

Figure \ref{fig2} shows an exemplary full-pattern Rietveld refinement (program FULLPROF) of the X-ray diffraction pattern for CeCo$_{0.6}$Fe$_{0.4}$Ge$_3$ ($x = 0.4$) sample. The lattice parameters are listed in Table \ref{tab1}. The structural analysis has shown that all the samples are single phase.

\begin{table}[h]
\centering
\caption{\label{tab1} Lattice parameters of CeCo$_{1-x}$Fe$_x$Ge$_3$ samples.}
\def\arraystretch{1.5}%
\begin{tabular}{ccc}
\hline
\hline
$x$ (Fe) & $a$ {\AA}& $c$ {\AA}\\
\hline
0.0 & 4.318 & 9.829\\
0.1 & 4.318 & 9.846\\
0.3 & 4.319 & 9.873\\
0.4 & 4.320 & 9.883\\
0.5 & 4.320 & 9.897\\
\hline
\hline
\end{tabular}
\begin{tabular}{ccc}
\hline
\hline
$x$ (Fe) & $a$ {\AA}& $c$ {\AA}\\
\hline
0.6 & 4.321 & 9.904\\
0.7 & 4.323 & 9.916\\
0.8 & 4.325 & 9.925\\
0.9 & 4.327 & 9.936\\
1.0 & 4.329 & 9.943\\
\hline
\hline
\end{tabular}
\end{table}

\begin{figure}[t!]
\centering
\includegraphics[width = 0.75\columnwidth]{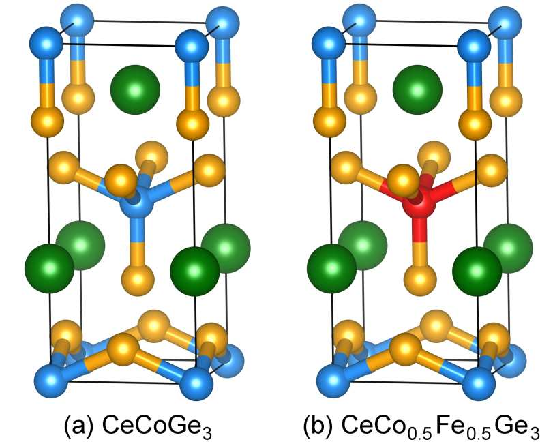}
\caption{\label{fig1} The crystal structure models of (a) CeCoGe$_3$ and (b) CeCo$_{0.5}$Fe$_{0.5}$Ge$_3$ compositions. The green balls represent Ce atoms, blue - Co, red - Fe, and yellow - Ge. These alloys crystallize in the tetragonal non-centrosymmetric BaNiSn$_3$-type structure.}
\end{figure}

\begin{figure}[t!]
\centering
\includegraphics[width = 0.9\columnwidth]{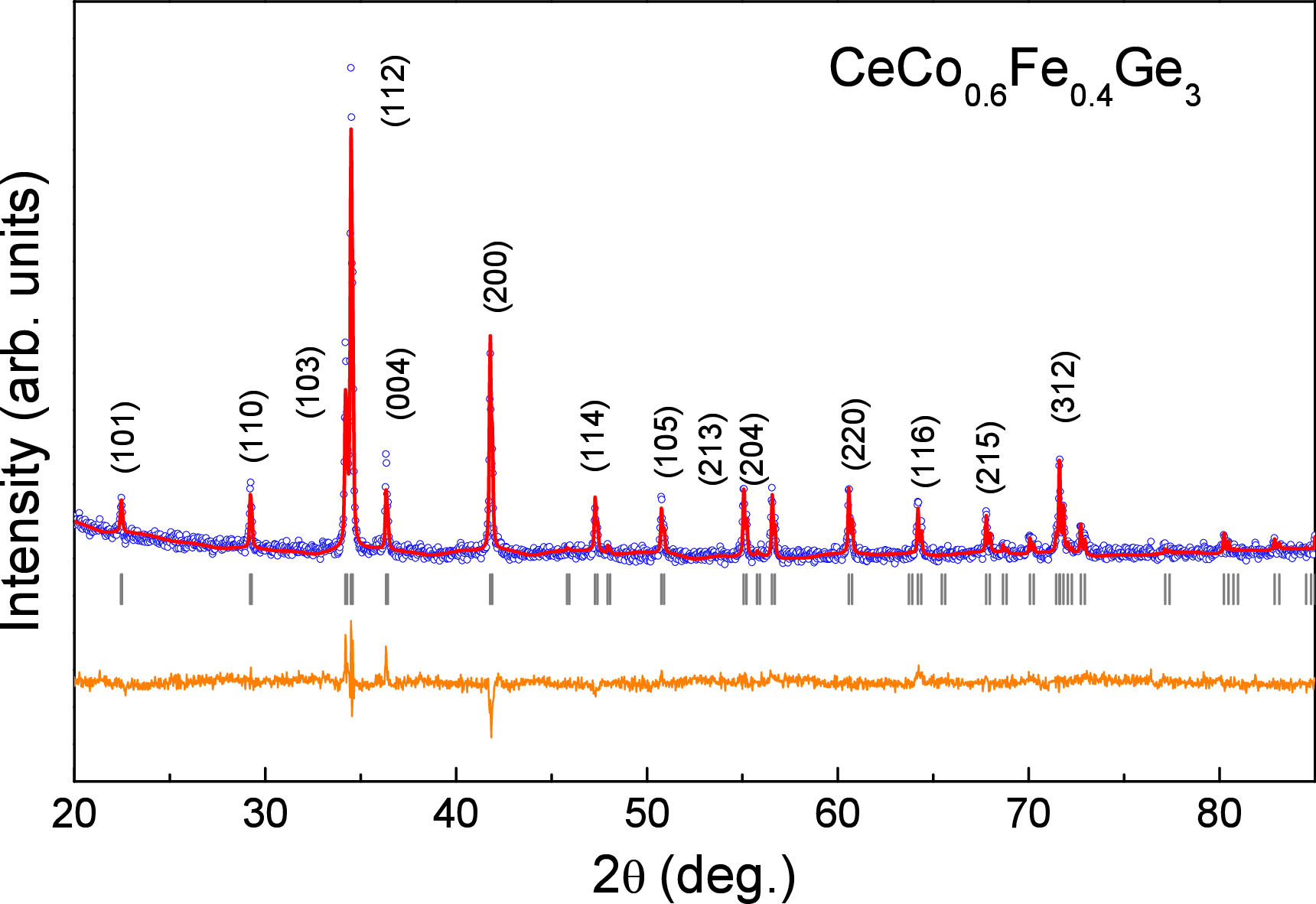}
\caption{\label{fig2} Representative XRD pattern of the CeCo$_{0.6}$Fe$_{0.4}$Ge$_3$ ($x = 0.4$) sample. The bottom solid line shows the difference between the measured and the calculated patterns. The vertical bars indicate the positions of the structural reflections. For the most pronounced peaks, Miller indices are shown.}
\end{figure}

\subsection{\label{mab}Calculations}

The electronic band structure calculations of the CeCoGe$_3$, CeCo$_{0.5}$Fe$_{0.5}$Ge$_3$, and CeFeGe$_3$ compositions were carried out using the full-potential local-orbital scheme (FPLO) \cite{koepernik1999full}. We used experimental lattice parameters. 
The intermediate composition CeCo$_{0.5}$Fe$_{0.5}$Ge$_3$ was modeled as an ordered compound with one Co and one Fe atom per unit cell (see Figure \ref{fig1}(b)). 
The calculations were performed in a full-relativistic mode within the generalized-gradient approximation (GGA) for exchange-correlation potential in the form of Perdew-Burke-Ernzerhof (PBE) \cite{perdew1996generalized} with the FPLO18.00-52 version of the code. We used 20$\times$20$\times$20 k-mesh and energy convergence criterion equal to 2.72$\times$10$^{-6}$ eV.

In a case of Ce-based compounds with an open Ce 4$f$ shell it is important to improve the local spin density approximation (LSDA) description by using for example the LSDA plus interaction term U (LSDA+U) approach. 
The LSDA+U adds an intra-atomic Hubbard U repulsion term in the energy functional and can significantly improve the results of LSDA.
A detailed insight into the LSDA+U methods is given for example in the article of Ylvisaker~\textit{et al.}~\cite{ylvisaker2009anisotropy}.
For the considered CeCo$_{1-x}$Fe$_x$Ge$_3$ compositions we used the fully localized limit (FLL) of the LSDA+U functional as introduced by Czy{\.z}yk and Sawatzky \cite{czyzyk1994local} sometimes referred also as the atomic limit (AL).
In our calculations the Hubbard U repulsion term introduced on Ce 4$f$ orbitals was gradually turned on with a step of 1 eV from 0 up to 6 eV.
Hubbard U for Ce equal to 6 eV has been calculated by Anisimov and Gunnarsson \cite{anisimov1991density}.

Based on the electronic band structure results we calculated the valence band X-ray photoelectron spectra.
To do that the partial densities of states (DOS) were convoluted by the Gaussian function with a full width at half maximum parameter $\delta$ equal to 0.3 eV, which was intended to imitate an experimental broadening coming from the instrumental resolution, lifetime of the hole states and thermal effects.
The partial DOSs were multiplied by the corresponding photoionization cross-sections \cite{yeh1985atomic}.
Previously we used the above XPS scheme on top of the LSDA+U (GGA+U) calculations for example in the study of $f$-electron compounds UGe$_2$ \cite{samsel2011electronic} and UCu$_2$Si$_2$~\cite{morkowski2011xray}.
For visualization of the crystal structures we used the VESTA code \cite{momma2008vesta}.

\section{\label{res}Results and discussion}
\subsection{\label{xps}X-ray photoelectron spectroscopy}
\begin{figure}[t!]
\centering
\includegraphics[width = 0.9\columnwidth]{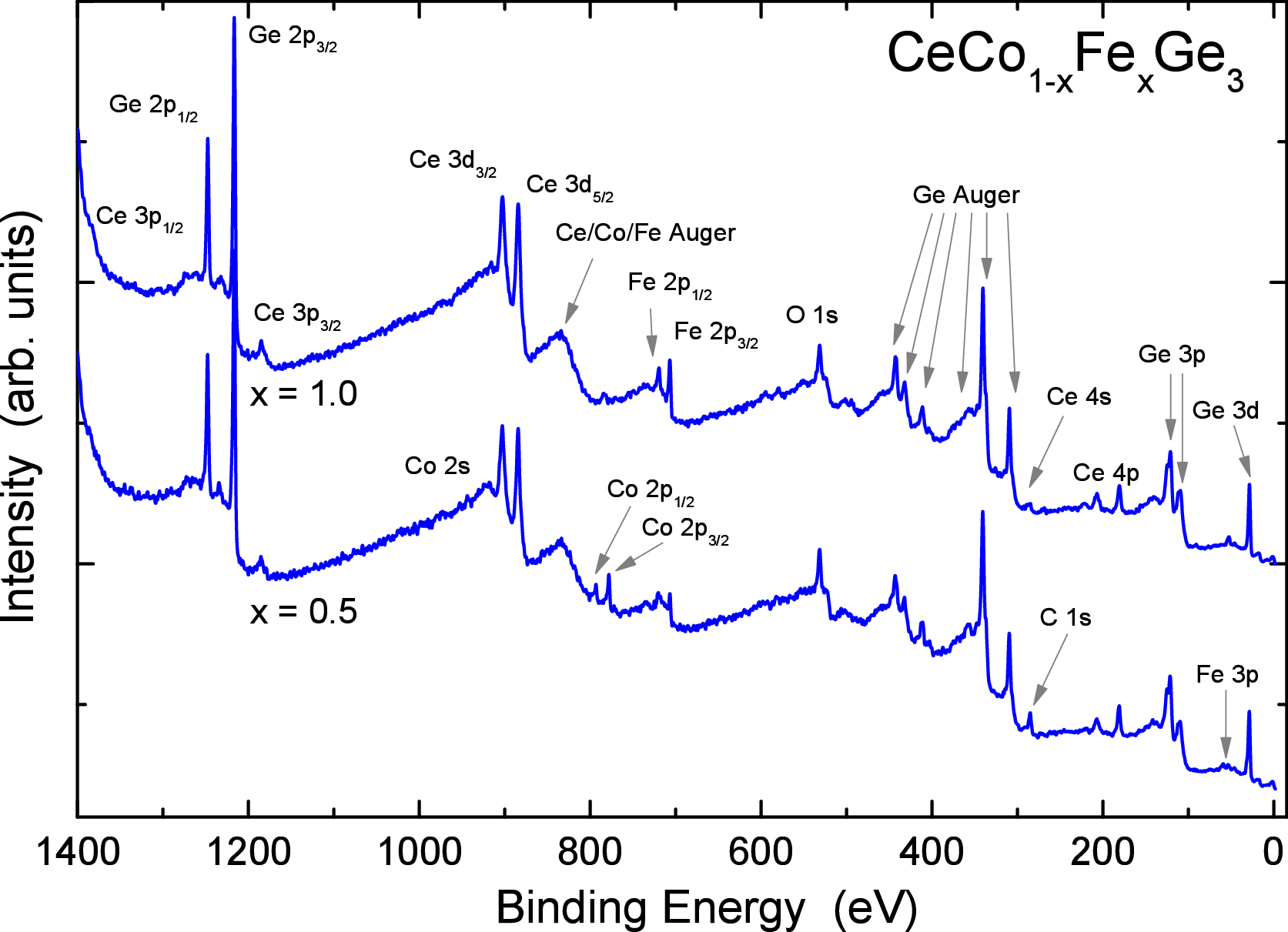}
\caption{\label{fig3} XPS spectrum collected at room temperature within the 0-1400 eV binding energy range for the example of the $x = 0.5$ and 1.0 samples of CeCo$_{1-x}$Fe$_x$Ge$_3$ series.}
\end{figure}

Figure \ref{fig3} shows exemplary X-ray photoelectron spectra of CeCo$_{1-x}$Fe$_x$Ge$_3$ alloys collected in a wide binding energy range up to 1400 eV.
A small intensity of the O $1s$ and C $1s$ peaks confirms high quality of the samples. 

\begin{figure}[t!]
\centering
\includegraphics[width = 0.9\columnwidth]{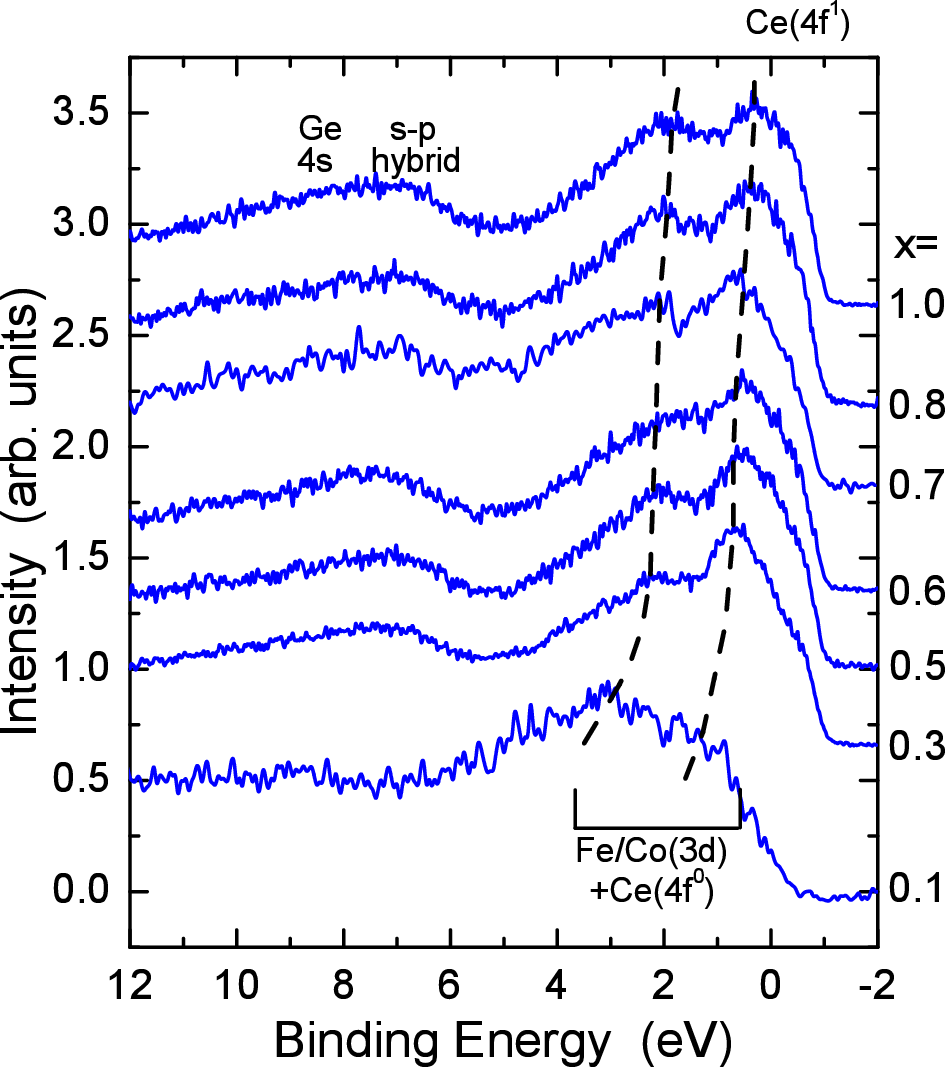}
\caption{\label{fig4} Valence XPS bands for the CeCo$_{1-x}$Fe$_x$Ge$_3$ alloys. The dashed lines illustrate the shift of the 3$d$ element peak and its typical satellite towards zero energy with the increase of the Fe content.}
\end{figure}

Figure \ref{fig4} shows the valence band spectra of the investigated alloys.
The main peaks are located between 0 and 4 eV and originate mostly from the Fe 3$d$, Co 3$d$, Ce 4$f^0$ and 4$f^1$ states.
Near the Fermi level ($E_{\rm F}$) there is a clearly visible hump that can be assigned to Ce $4f^1$. 
Usually, this multiplet structure can be well observed only in resonant XPS studies and is 
developed from the spin-orbit split Ce($4f_{7/2,\ 5/2}$) final states (280 meV)~\cite{kumigashira1996high,joyce1992temperature}.

\begin{figure}[t!]
\centering
\includegraphics[width = 0.8\columnwidth]{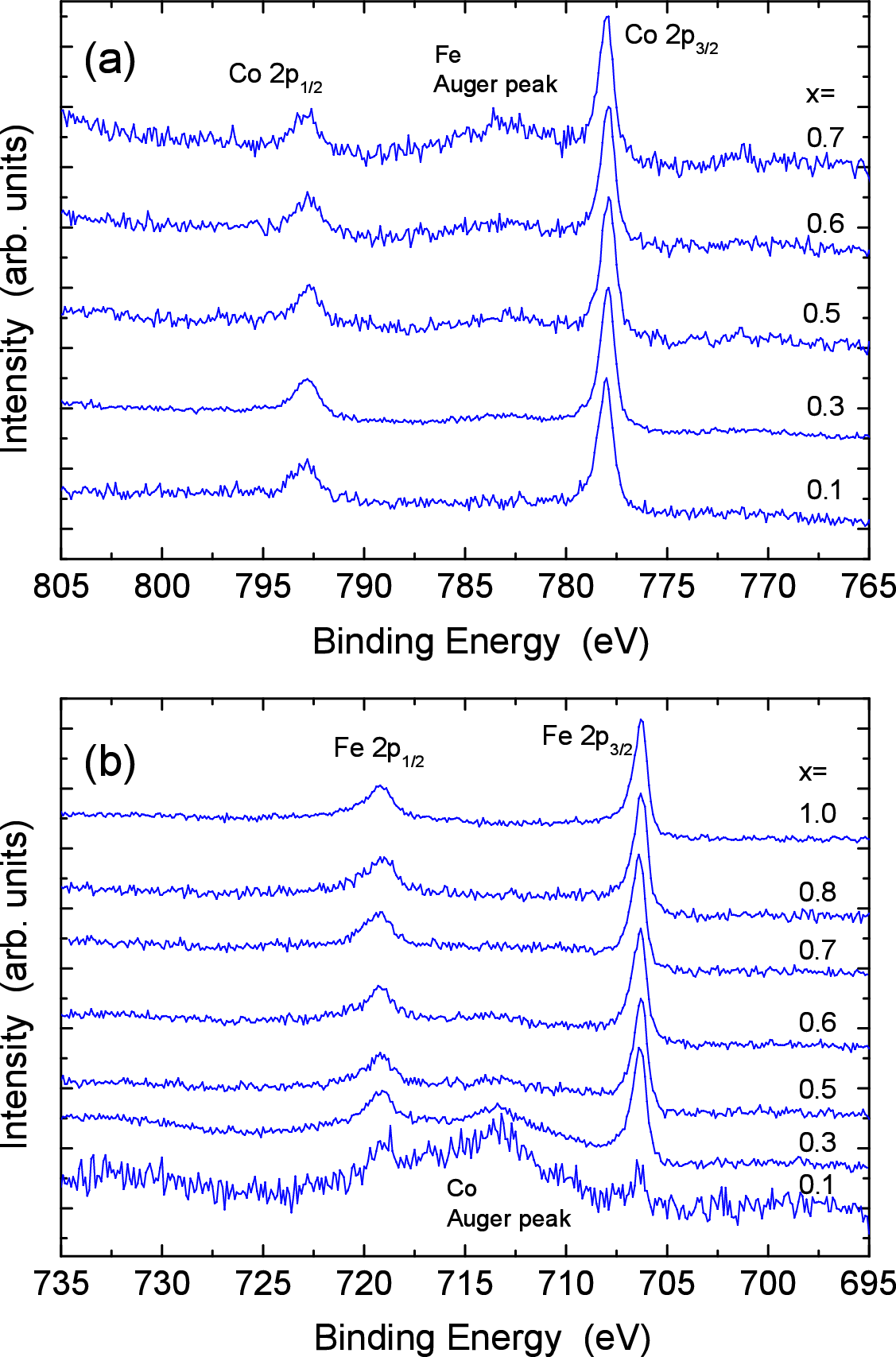}
\caption{\label{fig5}The Co $2p$ (a) and Fe $2p$ (b) XPS spectra for the CeCo$_{1-x}$Fe$_x$Ge$_3$ alloys.}
\end{figure}

In the Co $2p$ spectrum (Figure \ref{fig5}(a)), two distinct peaks at binding energies of 778 eV for Co $2p_{3/2}$ and 793 eV for $2p_{1/2}$ are observed for all samples.
The wide maximum between the spin-orbit split Co $2p$ peaks is ascribed to the Fe Auger intensity contribution, which is reasonable since this maximum is clearly increasing with the Fe content.
In the Fe $2p$ spectrum (Figure \ref{fig5}(b)), two peaks resulting from the spin-orbit splitting are observed at binding energies of 706.3 eV for Fe $2p_{3/2}$ and 719.2 eV for Fe $2p_{1/2}$ for all samples studied.
In analogy to the Co $2p$ case, the Auger peak is visible, this time due to the Co contribution.
The Fe $2p$ and Co $2p$ XPS spectra for CeCo$_{1-x}$Fe$_x$Ge$_3$ are all very similar to the spectra of pure metals.
We do not observe any additional satellite structures, which could be related to the charge-transfer effect and chemical shifts associated with oxides \cite{biesinger2011resolving}.
This suggests that Fe and Co may have a formal valency closer to 0 than 2+ \cite{mcleod2012effect}.
Valency close to 0 can be also supposed from the calculated charge being about 0.17 on the Co atoms for CeCoGe$_3$ and about $-0.03$ on the Fe atoms for CeFeGe$_3$ (see Table~\ref{tab_charge}).

\begin{figure}[t!]
\centering
\includegraphics[width = 0.7\columnwidth]{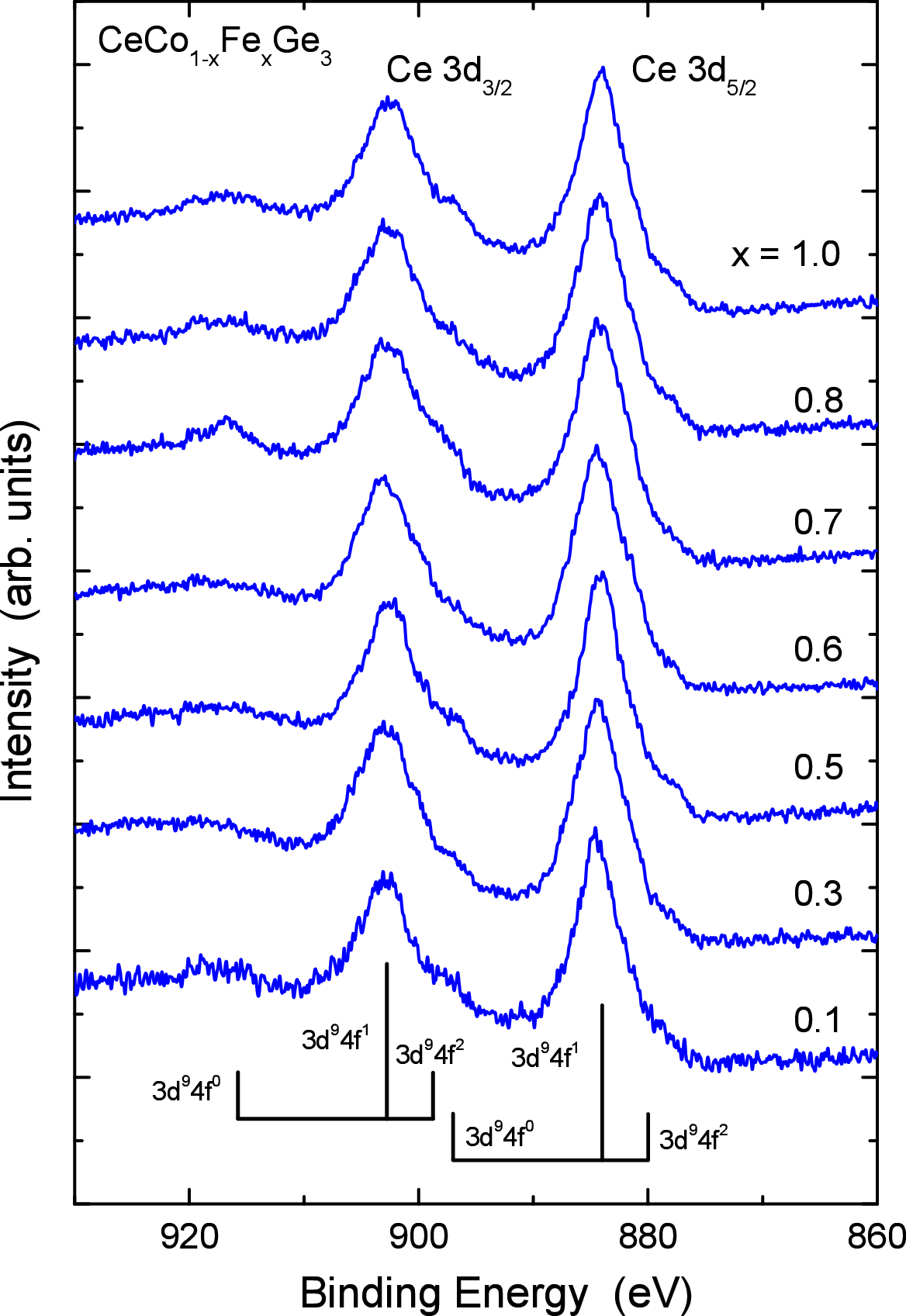}
\caption{\label{fig6} The Ce 3$d$ XPS spectra of the CeCo$_{1-x}$Fe$_x$Ge$_3$ alloys. The $f^n$ ($n = 0,\ 1,\ 2$) determines the final states related to the main peaks and the satellites. The Ce $3d_{5/2,\ 3/2}$ bands result from the spin-orbit splitting.}
\end{figure}

\begin{figure}[t!]
\centering
\includegraphics[width = 0.9\columnwidth]{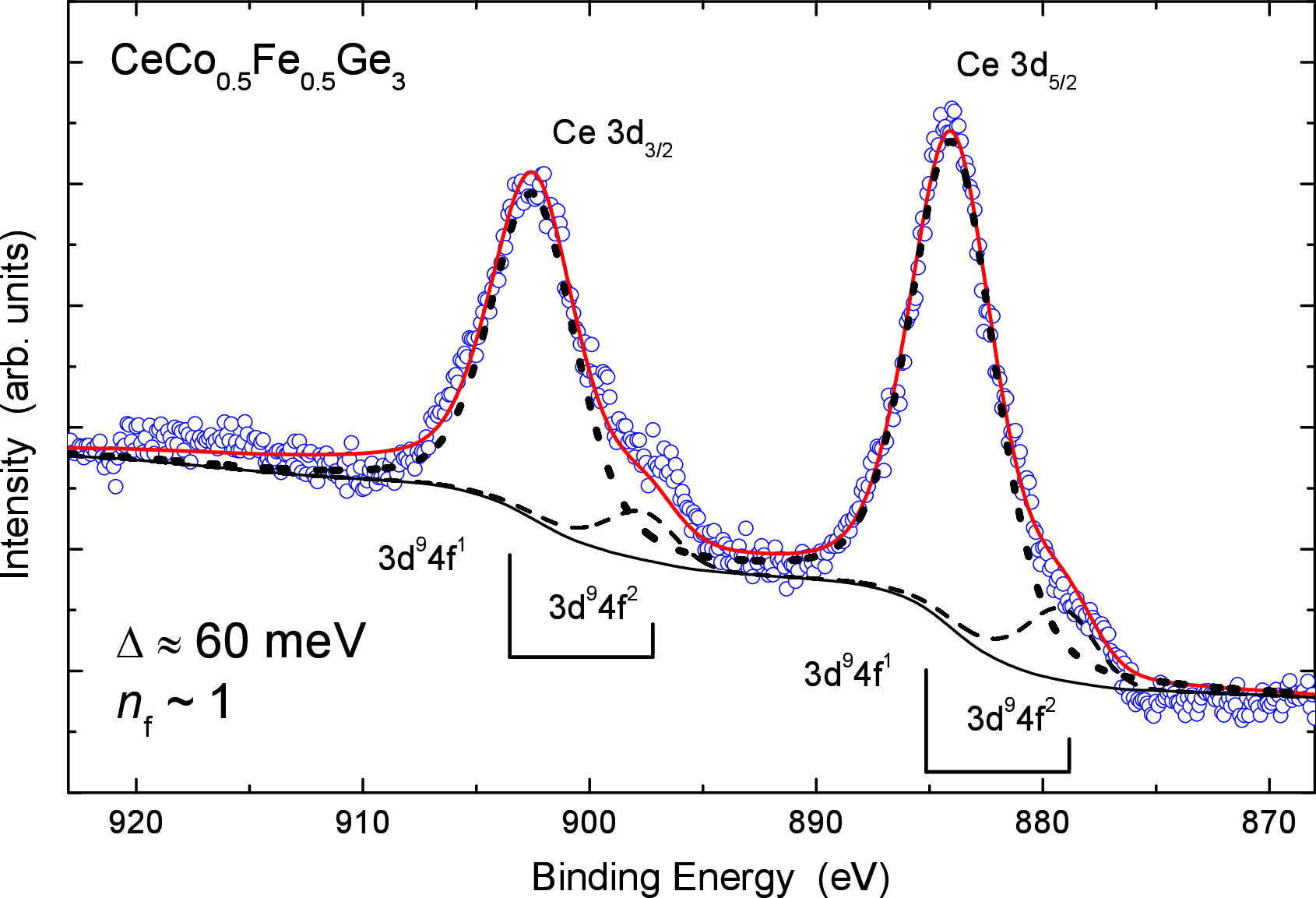}
\caption{\label{fig7} The exemplary analysis of the Ce 3$d$ XPS spectrum for CeCo$_{0.5}$Fe$_{0.5}$Ge$_3$. Open circles correspond to the experimental spectrum and the continuous curves to the fitting results (see text).}
\end{figure}

Figure \ref{fig6} shows the Ce 3$d$ core level spectra of the CeCo$_{1-x}$Fe$_x$Ge$_3$ alloys.
For all samples the contributions of the final states $4f^1$ and $4f^2$ are clearly observed, with the spin-orbit splitting of 18.8 eV.
The $f^0$ component is small indicating valence close to $3+$ (n$\sim$1) typical for magnetic Ce systems and Ce-based heavy fermion systems \cite{tolinski2011x,synoradzki2014x,tolinski2017influence,slebarski2009kondo}.
The first-principles calculations also indicate the occupancy for orbital 4$f$ is close to 1 (values 1.0 $\pm$ 0.1), see Table \ref{tab_orbital_occupation}.
The fluctuations of the valence seem negligible.
The Ce 3$d$ spectra have been analyzed on the base of the theoretical model proposed by Gunnarsson and Sch{\"o}nhammer \cite{gunnarsson1983gunnarsson}.
Within this model the $f^2$ peak intensity reflects the hybridization strength, therefore the energy coupling $\Delta$ between the $f$ electrons and conduction states can be derived from the intensity ratio $r_2 = I(f^2)/[I(f^1)+I(f^2)]$.
The 4$f$-occupancy n$_f$ can be estimated from the ratio $r_0 = I(f^0)/[I(f^0) + I(f^1) + I(f^2)]$, because the $f^0$ satellite provides information on the $f$-electron counts.
The value of n$_f \sim 1$ is attributed to the stable Ce$^{3+}$ state, in the case of the fluctuating valence character of the Ce ions n$_f < 1$.

Figure \ref{fig7} presents an example of the spectrum decomposition for the CeCo$_{0.5}$Fe$_{0.5}$Ge$_3$. 
The Shirley method \cite{shirley1975sensitivity} was used to subtract the background.
The analysis with Gunnarsson and Sch{\"o}nhammer model have shown that the $f$-occupancy is close to 1 for the entire range of the Fe substitution, which is reasonable taking into account that both for the magnetically ordered alloys and the heavy fermions a localized character of the $f$ states is typical.
The estimated hybridization parameter, $\Delta$, varies non-systematically between 30-70 meV.
The localization of the $f$ states resulting from the valence band (Fig. \ref{fig4}) and the Ce $3d$ spectra (Fig. \ref{fig7}) supports the possibility of QCP appearence in the CeCo$_{1-x}$Fe$_x$Ge$_3$ series.

\subsection{\label{abi}First-principles calculations}

\subsubsection{X-ray Photoelectron Spectra}

\begin{figure}[t!]
\centering
\includegraphics[width = \columnwidth]{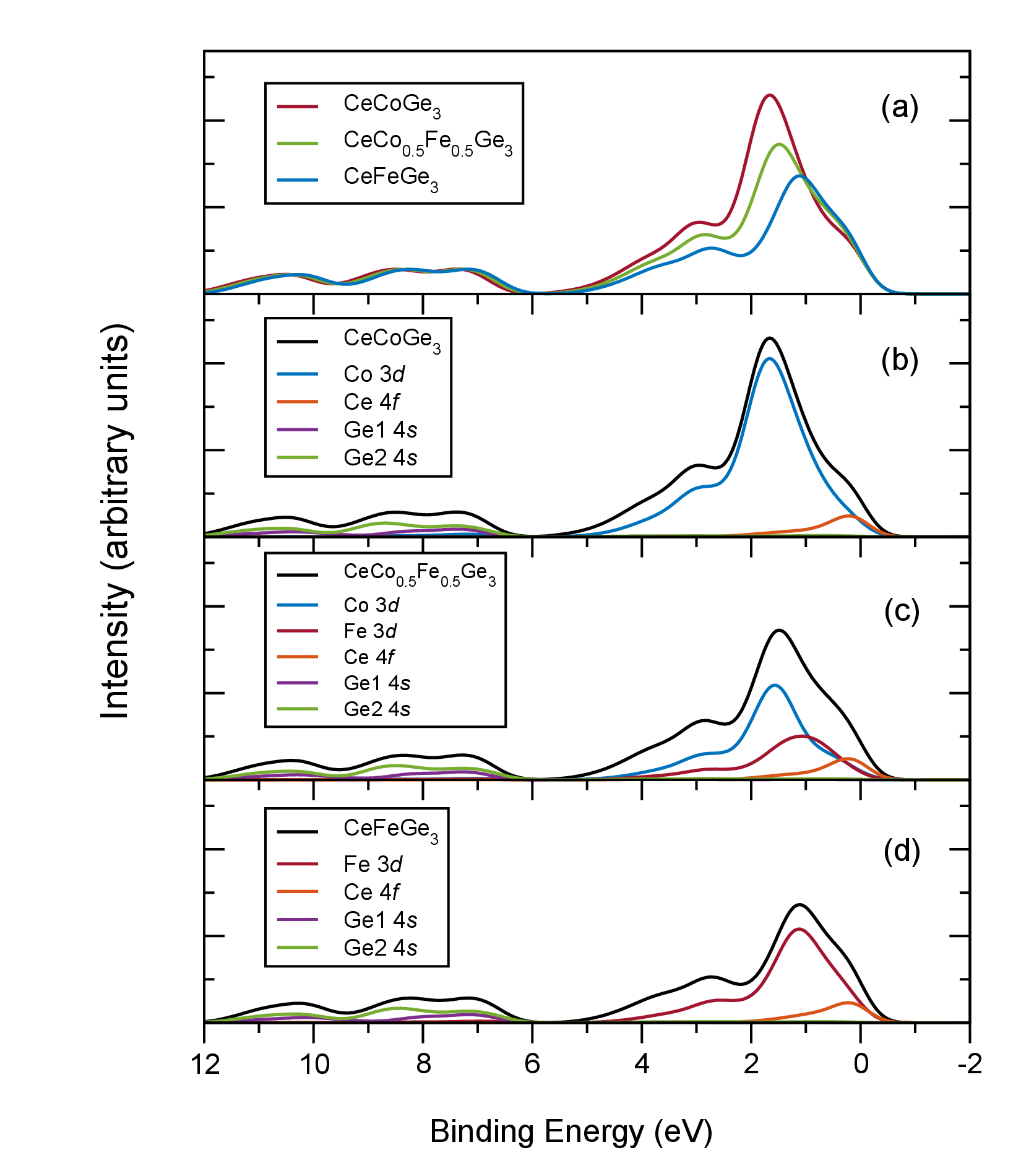}
\caption{\label{fig8} 
The valence band X-ray photoelectron spectra as calculated with the PBE (U = 0, $\delta = 0.3$ eV). 
Calculations were done within the FPLO18 code using the PBE functional and treating the relativistic effects in a full 4-component formalism (including spin-orbit coupling).
(a) Comparison of the results for CeCoGe$_3$, CeCo$_{0.5}$Fe$_{0.5}$Ge$_3$, and CeFeGe$_3$. (b, c, d) The most significant contributions from individual orbitals compared to the total XPS spectra.}
\end{figure}

The main aim of our first-principles study is to interpret the measured XPS valence band spectra.
For this purpose we solve self-consistently the models for the most characteristic compositions: CeCoGe$_3$, CeCo$_{0.5}$Fe$_{0.5}$Ge$_3$, and CeFeGe$_3$ and evaluate the corresponding densities of states and X-ray photoelectron spectra. 
A presence of the 4$f$ element like Ce in considered system motivates us to go beyond the LSDA formalism and apply the LSDA plus interaction term U (LSDA+U) approach. 
Because in our case the interaction term U applied to Ce 4$f$ orbitals does not affect noticeably the part of the valence band located below the $E_{\rm F}$, the XPS spectra calculated with and without U look nearly identical. 
Thus, for a clarity of presentation, in Fig.~\ref{fig8} we show only the spectra calculated without U.

The comparison of the calculated XPS for CeCoGe$_3$, CeCo$_{0.5}$Fe$_{0.5}$Ge$_3$, and CeFeGe$_3$, as presented in Fig. \ref{fig8}(a), reveals two characteristic features of how the spectra evolve with increasing of the Fe concentration in the samples.
First observation is a shift of the 3$d$-element contribution towards zero energy level and the second observation is a decrease of intensity of the 3$d$ contribution with $x$.
The explanations of these behaviours are (1) a reduction of the number of occupied 3$d$ states when alloying Fe for Co (as the neutral Fe and Co atoms have 26 and 27 electrons, respectively) and (2) a large reduction of the Fe 3$d$ photoionization cross-section in respect to Co 3$d$ cross-section (0.0022 and 0.0037, respectively) \cite{yeh1985atomic}.
The main features of the experimental XPS spectra, and the evolution of the spectra with Fe alloying, are correctly reproduced by theory.
The most significant difference between the experimental and theoretical XPS is the underestimation of the maxima observed at about 3.0 eV.
We ascribe it to the $4f^0$ contribution, which overlaps with the Ni/Co $3d$ peaks (see Fig. \ref{fig4}).
Theoretically, only Anderson-type dynamical models could provide more details on the $f$-contributions.
Figures \ref{fig8}(b-d) present the most significant contributions from individual orbitals to the XPS spectra of CeCoGe$_3$, CeCo$_{0.5}$Fe$_{0.5}$Ge$_3$, and CeFeGe$_3$.
The region between 0 and 5 eV is formed mainly by the 3$d$ shares of Fe and Co with minor contribution of the Ce 4$f$ states around 0 eV.
The second observed band, located between 6 and 12 eV, consists mainly of the Ge $4s$ states.

\subsubsection{Densities of States}

\begin{figure}[t!]
\centering
\includegraphics[width = \columnwidth]{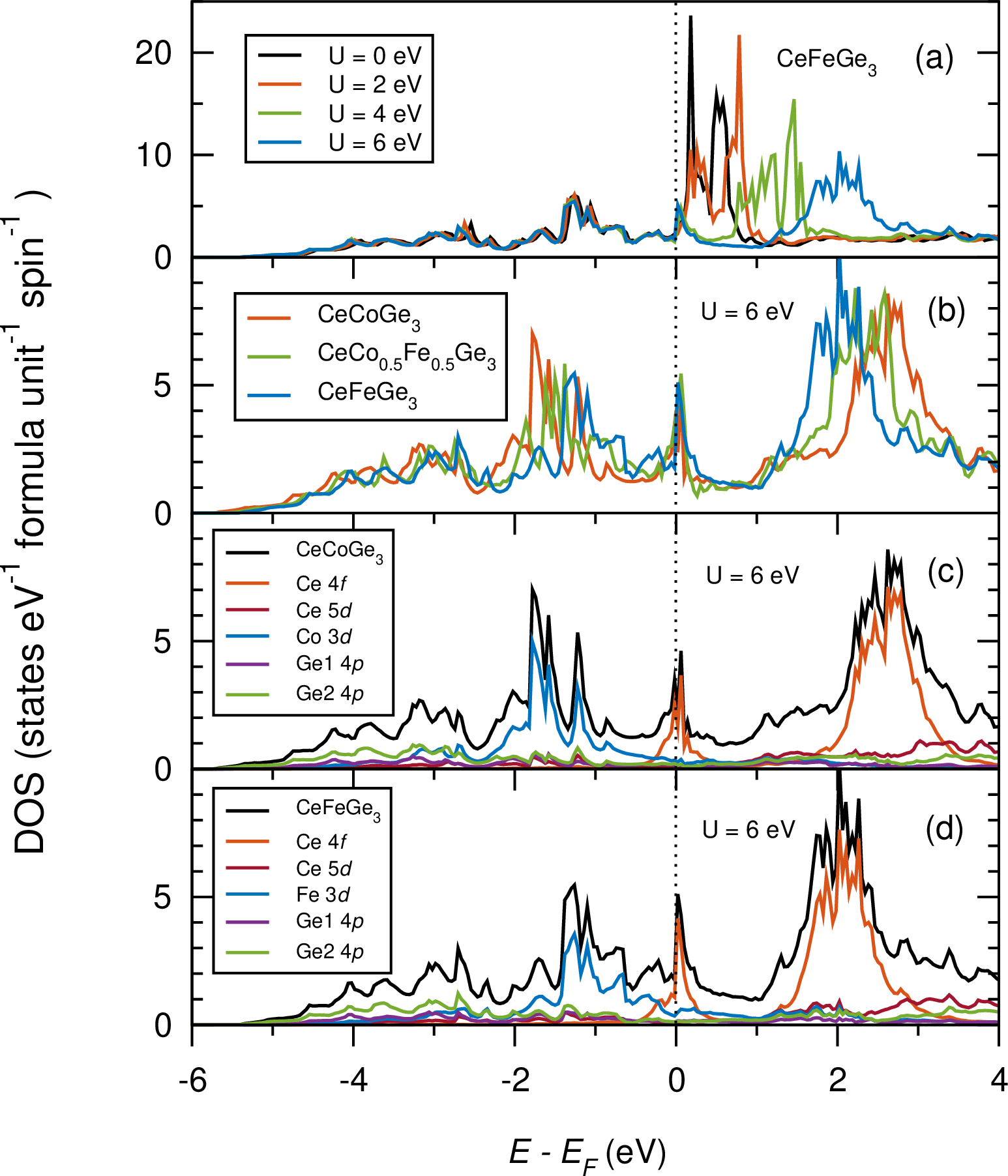}
\caption{\label{fig9} 
Densities of states (DOS) calculated with taking into account an intra-atomic Hubbard U repulsion term U(Ce 4$f$) equal to 6 eV.
Calculations were done within the FPLO18 code using the PBE functional and treating the relativistic effects in a full 4-component formalism (including spin-orbit coupling).
(a) Comparison of DOS for the increasing value of U for the representative case of CeFeGe$_3$. 
(b) Comparison of the results for CeCoGe$_3$, CeCo$_{0.5}$Fe$_{0.5}$Ge$_3$, and CeFeGe$_3$. 
(c, d) The most significant contributions from individual orbitals compared to the total DOS.
}
\end{figure}

The calculated XPS spectra are based on the partial DOSs as shown in Fig.~\ref{fig9}.
The presented DOSs plots cover both the occupied valence band region below $E_{\rm F}$ and the unoccupied region above $E_{\rm F}$ -- not probed in the XPS measurements. 
The region above $E_{\rm F}$ consists mainly of a contribution from the Ce 4$f$ states.
Without taking into account the intra-atomic Hubbard U repulsion on Ce 4$f$ electrons (U = 0), 
the full relativistic treatment of electrons (including spin-orbit coupling) leads to a picture in which the Ce 4$f$ states are separated into two 4$f_{5/2}$ and 4$f_{7/2}$ peaks located in a region between 0 and 1 eV.
The application of the Hubbard U results in further separation of the Ce 4$f$ states into a part located close to $E_{\rm F}$ and second one pushed towards unoccupied states (see Fig. \ref{fig9}(a)).
Similar conclusion regarding the effect of spin-orbit coupling and Hubbard U repulsion on electronic structure of CeCoGe$_3$ can be found in work of Jeong~\cite{jeong2007electronic}.
Whereas, the DOS of CeFeGe$_3$ calculated with U$_{4f}$ = 6 eV resembles in main characteristics the results of Chigo-Anota~\textit{et al.} \cite{chigo2006lsda} obtained with the LMTO-ASA method.
However, in comparison to the results of Ref. \cite{chigo2006lsda} the DOS presented in Fig. \ref{fig9}(a) resolves additionally the 4$f$ peak close to $E_{\rm F}$.
Similar picture of DOS on non-interacting 4$f$-3$d$ orbitals has been found also, for example, for La$_2$Co$_7$~\cite{kuzmin2015magnetic}.
Despite the fact that Chigo-Anota~\textit{et al.}~\cite{chigo2006lsda} have used in their model of CeFeGe$_3$ an additional Hubbard U repulsion on Fe 3$d$ orbitals (U$_{Fe}$ = 2.3 eV), we decided to consider the Hubbard U interactions only for the Ce 4$f$ orbitals.
We have tested that the effect of the Hubbard U on the Fe/Co 3$d$ orbitals is weak, thus the U$_{3d}$ corrections can be omitted. 
%
%
%
%
Figure \ref{fig9}(b) presents how the valence band of the CeCo$_{1-x}$Fe$_x$Ge$_3$ system evolves with increasing the Fe concentration. 
It stays in a natural agreement with the previously described evolution of the calculated XPS, where the main feature is the shift of the 3$d$ band towards $E_{\rm F}$ with the increase of Fe concentration. 
Similar shift is also observed for the part of Ce 4$f$ band located around 2 eV.
For each of the considered compositions a localized part of the Ce 4$f$ band, induced by the applied intra-atomic Hubbard U repulsion, is also observed at the $E_{\rm F}$.

%
Figures \ref{fig9}(c) and \ref{fig9}(d) present the most significant contributions from individual orbitals together with the total DOS. 
For a clarity of the picture, only the selected contributions of individual orbitals (Co/Fe 3$d$, Ce 4$f$, Ce 5$d$, and Ge 4$p$) are presented.
Another important orbital, Ge 4$s$, is not shown, as its main part is located below the energy range covered by the plot.
The Ge 4$s$ orbital hybridize with Ge 4$p$ and take a part in formation the (Fe/Co)-Ge bonds.
Below $E_{\rm F}$ one can observe a hybridization between Fe/Co 3$d$ and Ge $4p$ orbitals providing formation of 3$d$-4$p$ bonds between the nearest neighbors Fe/Co and Ge (see Fig.~\ref{fig1}). 
The detailed analysis of hybridization between orbitals is possible for example by applying techniques presented by Goraus and \'{S}lebarski~\cite{goraus2011onsite}, however it is beyond the scope of this work.
The main contribution above $E_{\rm F}$ comes from the Ce 4$f$ states.
Differences observed between the total DOS and contributions from Ce 4$f$ states at, for example, 1 and 4 eV, come from the character of the total DOS at these energy values.
At 1 eV the total DOS is a combination of several contributions from which the largest ones come from the Ge 5$d$, Fe 3$d$, Ge2 4$p$, and Ge1 4$p$ orbitals.
At 4 eV the total DOS is mainly composed of Ce 5$d$ and Ge2 4$p$ contributions.

%
CeCoGe$_3$ is an antiferromagnet with the electronic specific heat coefficient $\gamma$ equal to 111~mJ/(mol\,K$^2$) if determined at low temperatures below the antiferromagnetic ordering temperatures~\cite{pecharskyunusual1993,jeong2007electronic}. 
CeFeGe$_3$ is paramagnetic and it is a moderate heavy fermion compound with electronic specific heat coefficient equal to 150~mJ/(mol\,K$^2$)~\cite{yamamotonew1994, yamamotocefege1995}.
The calculated values of DOS(\textit{E}$_\mathrm{F}$) for CeCoGe$_3$ and CeFeGe$_3$ are 11.96 (4.45) and 5.9 (4.78)~states/(eV\,f.u.), respectively, whereas in parentheses are shown the results obtained with U = 0.
The values of DOS(\textit{E}$_\mathrm{F}$) for U = 6 eV correspond to the $\gamma$ equal to 28.2 and 13.91~mJ/(mol\,K$^2$) for CeCoGe$_3$ and CeFeGe$_3$, respectively.
The calculated values of DOS(\textit{E}$_\mathrm{F}$) consist mainly of the contributions from Ce 4$f$ and Fe/Co 3$d$ electrons, see Fig.~\ref{fig9}(c) and (d).
%
%
Jeong~\cite{jeong2007electronic} has attributed the significant underestimation of the theoretical $\gamma$ of CeCoGe$_3$ to the formation of quasi-particles between the local $f$ electrons and conduction electrons.

\subsubsection{Charge Analysis}

\begin{table}[ht!]
\centering
\caption{\label{tab_charge} Excess electron number (resultant charge) for CeCoGe$_3$ and CeFeGe$_3$ compounds, as calculated with and without taking into account the intra-atomic Hubbard U repulsion (in eV) on Ce 4$f$ orbitals. 
Calculations were done within the FPLO18 code using the PBE functional and treating the relativistic effects in a full 4-component formalism (including spin-orbit coupling).
}
\def\arraystretch{1.5}%
\begin{tabular}{l|cc|cc}
\hline
\hline
	&CeCoGe$_3$	&CeFeGe$_3$	&CeCoGe$_3$ 	&CeFeGe$_3$	\\
	&U = 0		&U = 0		&U = 6	&U = 6\\
\hline
Ce	&-1.16		&-1.16		&-1.23	&-1.24 \\
Co/Fe	&0.18		&-0.01		&0.17	&-0.03 \\
Ge(1)	&0.42		&0.49		&0.47	&0.55 \\
Ge(2)	&0.28		&0.34		&0.30	&0.36 \\
Ge(3)	&0.28		&0.34		&0.30	&0.36 \\
sum	&0.00		&0.00		&0.00	&0.00 \\
\hline
\hline
\end{tabular}
\end{table}

%
%
Table~\ref{tab_charge} presents the excess electron numbers (resultant charges) for the CeCoGe$_3$ and CeFeGe$_3$ compounds calculated with and without taking into account the intra-atomic Hubbard U repulsion on Ce 4$f$ orbitals.
The resultant charge on Ce (for U = 0) is about -1.16, which is much higher than -2.4 obtained by Chigo-Anota \textit{et al.} \cite{chigo2006lsda}. 
However, the accuracy of the latter result is limited by not taking into account the full potential and spin-orbit coupling.
The charge from Ce is transfered mainly to the Ge sites contributing to the formation of ionic bonds between the nearest neighbors Ce and Ge (see Fig.~\ref{fig1}).
The resultant charge for Co is equal to 0.18 and for Fe equals -0.01, which indicates a nearly neutral character of Fe and weak charge polarization for Co.
After introduction of the intra-atomic Hubbard U repulsion on Ce 4$f$ orbitals, a slightly more charge is transfered from Ce to other sites, in result giving a value of about -1.23/-1.24 electrons on Ce (instead of -1.16) and increasing the charge number on Ge sites.

\begin{table}[ht!]
\centering
\caption{\label{tab_orbital_occupation} Population analysis for CeCoGe$_3$ and CeFeGe$_3$ compounds as calculated with and without taking into account the intra-atomic Hubbard U repulsion (in eV) on Ce 4$f$ orbitals.
Calculations were done within the FPLO18 code using the PBE functional and treating the relativistic effects in a full 4-component formalism (including spin-orbit coupling).
Several nearly empty valence orbitals are not presented, although they were considered in calculations.}
\def\arraystretch{1.5}%
\begin{tabular}{cccccccc}
\hline
\hline
   & U  & site& 5$p$  &  6$s$ &  5$d$ &  6$p$ &  4$f$\\
CeCoGe$_3$ &  0 &  Ce &  5.84 &  0.19 &  1.46 &  0.13 &  1.15\\
CeCoGe$_3$ &  6 &  Ce &  5.82 &  0.19 &  1.59 &  0.14 &  0.98\\
CeFeGe$_3$ &  0 &  Ce &  5.84 &  0.19 &  1.48 &  0.14 &  1.12\\
CeFeGe$_3$ &  6 &  Ce &  5.82 &  0.20 &  1.64 &  0.15 &  0.91\\
\hline
    & U  &  site& 4$s$ &  3$d$ &  4$d$ &  4$p$\\
CeCoGe$_3$ &  0 &  Co &  0.52 &  7.91 &  0.19 &  0.55\\
CeCoGe$_3$ &  6 &  Co &  0.52 &  7.91 &  0.19 &  0.53\\
CeFeGe$_3$ &  0 &  Fe &  0.48 &  6.90 &  0.17 &  0.40\\
CeFeGe$_3$ &  6 &  Fe &  0.48 &  6.93 &  0.16 &  0.39\\
\hline
    & U  &  site& 4$s$ &  4$p$ & 4$d$\\
CeCoGe$_3$ &  0 &  Ge1 & 1.59 &  2.66 & 0.16\\ 
CeCoGe$_3$ &  6 &  Ge1 & 1.60 &  2.71 & 0.16\\ 
CeFeGe$_3$ &  0 &  Ge1 & 1.61 &  2.71 & 0.16\\
CeFeGe$_3$ &  6 &  Ge1 & 1.62 &  2.76 & 0.16\\
\hline
    & U  &  site& 4$s$ &  4$p$ & 4$d$\\
CeCoGe$_3$ &  0 &  Ge2 & 1.60 &  2.53 & 0.17\\ 
CeCoGe$_3$ &  6 &  Ge2 & 1.60 &  2.55 & 0.17\\
CeFeGe$_3$ &  0 &  Ge2 & 1.61 &  2.58 & 0.17\\
CeFeGe$_3$ &  6 &  Ge2 & 1.61 &  2.60 & 0.17\\
\hline
\hline
\end{tabular}
\end{table}

%
%
More details regarding occupation of orbitals in CeCoGe$_3$ and CeFeGe$_3$ can be found in Table~\ref{tab_orbital_occupation}.
For CeCoGe$_3$ (U = 0),
after transferring of about 1.16 electrons (total charge on site) from the neutral Ce to the Co and Ge sites,
the electronic configuration on the Ce site is 5$p^{5.84}$ 6$s^{0.19}$ 5$d^{1.46}$ 6$p^{0.13}$ 4$f^{1.15}$.
We observe a small depopulation of 5$p$ orbitals and a small occupation of 6$s$ and 6$p$ orbitals.
The noticeable occupations of 1.15 and 1.46 are found for 4$f$ and 5$d$ orbitals, respectively.
In respect to a ground state electronic configuration of a neutral Ce atom (4$f^1$ 5$d^1$ 6$s^2$), we observe a severe depopulation of the Ce 6$s$ orbital, which looses its electrons for two other orbitals on the Ce site, as well as for other sites.

%
In Table~\ref{tab_orbital_occupation} are also presented the results of the orbital occupation for Fe and Co.
For CeCoGe$_3$ (U = 0), after attraction of about 0.18 electrons, 
the electronic configuration on the Co site is 4$s^{0.52}$ 3$d^{7.91}$ 4$d^{0.19}$ 4$p^{0.55}$.
In respect to a ground state electronic configuration of a neutral Co atom (3$d^7$ 4$s^2$), we observe depopulation of the 4$s$ orbital together with a simultaneous increase in occupation of the 3$d$ orbital.
Additionally, the 4$d$ and 4$p$ orbitals become partially occupied.
In transition from CeCoGe$_3$ to CeFeGe$_3$, the main difference in electronic configuration of Co and Fe is in occupation of the 3$d$ orbital (7.91 \textit{versus} 6.90), which comes from a difference in nuclear charge between Co and Fe.

%
The electronic configurations on the Ge1 and Ge2 sites in CeCoGe$_3$ (U = 0) are similar and for Ge1 it is 4$s^{1.59}$ 4$p^{2.66}$ 4$d^{0.16}$ (see Table~\ref{tab_orbital_occupation}).
The already mentioned charge attracted from Ce sites has been located on Ge 4$p$ orbital, and the Ge 4$p$ orbitals hybridize with Fe/Co 3$d$ orbitals, which also have attracted charge (see Figs. \ref{fig9}(c, d) \ref{fig9}d, and Table~\ref{tab_orbital_occupation}).
In respect to a ground state electronic configuration of a neutral Ge atom (3$d^{10}$ 4$s^2$ 4$p^2$), for Ge1 we observe accumulation of charge on 4$p$ and 4$d$ orbitals and partial depopulation of 4$s$.

%
Concluding, the bondings in CeCoGe$_3$ and CeFeGe$_3$ result from an interplay of several atomic orbitals, which are mainly Ce 5$d$, Fe/Co 3$d$, and Ge 4$p$ and 4$s$.
These orbitals gain charge at the expense of the $s$-type valence orbitals of all site's types.
The occupied Ce 4$f$ states, located close to the $E_{\rm F}$, do not take a part in the formation of bondings.

\section{Conclusions}
The presented results show that the substitution of Fe for Co in the series CeCo$_{1-x}$Fe$_x$Ge$_3$ influence significantly the electronic structure of the alloys, which can be summarized as follows:
\begin{labeling}{(iii)}
\item [(i)]The 3$d$ contribution within the valence band shifts quickly towards Fermi level with the increase of the Fe content.
\item [(ii)]Both the hump at $E_{\rm F}$ due to the Ce $f^2$ contribution at the valence band as well as the analysis of the satellites for the Ce 3$d$ spectrum indicate localization of the $f$ states for all $x$ values, i.e. the valence is close to $3+$ and the hybridization of the 4$f$ electrons with the conduction electrons is weak.
\item [(iii)] For the considered compositions the calculations indicate electronic state close to Ce $f^1$.

\item [(iv)]The calculations explain the evolution of the measured CeCo$_{1-x}$Fe$_x$Ge$_3$ XPS spectra with $x$ as resulting from the decrease of the number of electrons in the system and the reduction of the 3$d$  photoionization cross-sections. 

\item [(v)] Density of states and charge analysis indicate that the bondings in CeCoGe$_3$ and CeFeGe$_3$ are formed mainly by Ce 5$d$, Fe/Co 3$d$, and Ge 4$p$ and 4$s$ orbitals.

\item [(vi)] The calculations indicate that charge transfer occurs mainly from Ce towards Ge sites.

\end{labeling}

\section*{Acknowledgments}
MW acknowledges the financial support from the Foundation of Polish Science grant HOMING. The HOMING programme is co-financed by the European Union under the European Regional Development Fund. Part of the computations was performed on the resources provided by the Pozna{\'n} Supercomputing and Networking Center (PSNC).

\bibliography{bib}

\end{sloppypar}
\end{document}